**Wafer-scale monolithic integration of Ce:YIG films and magneto-optical isolators on silicon**

*Tianchi Zhang[1,2], Yucong Yang[3], Weihao Yang[4], JieJun Su[1,2], Tianyi Ma[1,2], Xuan Zhao[1,2], Junxian Wang[1,2], Di Wu[1,2], Zhenyuan Ren[1,2], Yi Shuai[1,2], Zixuan Wei[1,2], and Lei Bi[1,2]**

1.National Engineering Research Centre of Electromagnetic Radiation Control Materials, University of Electronic Science and Technology of China, Chengdu 611731, China

2.Key Laboratory of Multi-spectral Absorbing Materials and Structures of Ministry of Education, University of Electronic Science and Technology of China, Chengdu 611731, China

3.Institute of Optics and Electronics, Chinese Academy of Sciences, Chengdu, Sichuan 610209, China

4.HK Institute of Quantum Science and Technology, University of Hong Kong, Hong Kong 999077, China

E-mail: bilei@uestc.edu.cn

**Abstract: Silicon integrated cerium doped yttrium iron garnet (Ce:YIG) thin films are promising candidates for integrated nonreciprocal photonic devices, cryogenic photonic modulators and optical computing applications. However, previously reported Ce:YIG thin film on silicon is limited to milimeter sizes. Wafer-scale integration and non-destructive characterization of high quality Ce:YIG thin films on silicon has been elusive. Here, we report growth of 4-inch wafer-scale Ce:YIG thin films on silicon substrates by radio-frequency magnetron sputtering. Strong Faraday effect of 2318±102 deg/cm, low propagation loss of 80±11 dB/cm and excellent thickness uniformity of 3.5% is demonstrated across the 4-inch silicon wafer. Furthermore, a custom designed wafer-scale, non-destructive magneto-ellipsometry was established to characterize the film thickness, optical constants and magneto-optical constants across the wafer. Wafer-scale integration of ring resonator type magneto-optical isolators are also demonstrated. Our work demonstrates a step forward toward wafer-scale heterogeneous integration and characterization of magneto-optical thin films on silicon, providing material candidates for non-reciprocal photonic device arrays, magneto-optical in-memory computing networks and integrated magneto-optic magnetometers.**

## 1. Introduction

The rapid development of artificial intelligence leads to an increasing demand for photonic integrated circuits (PICs) for high speed data communication, optical computing and opticla sensing applications. Magneto-optical devices constitute essential components in photonic integrated circuits (PICs), serving key roles in optical isolation[1], nonreciprocal optical routing[2], in-memory optical computing[3], cryogenic optical modulation[4] and quantum information processing[5]. In recent years, silicon integrated magneto-optical (MO) cerium doped yttrium iron garnet (Ce:YIG) thin films have emerged as promising material candidate, thanks to its remarkable MO figure of merit (FoM) and strong Faraday rotation at optical communication wavelengths. Integrated optical isolators, circulators, switches and modulators have been demonstrated based on Ce:YIG on silicon, demonstrating high isolation ratio over 30 dB, low insertion loss less than 0.5 dB and compact device size less than 0.2 $mm^2$ that are comparable or even surpassing discrete devices[1, 6-14].

Successful development of above photonic devices require heterogeneous integration of high quality Ce:YIG thin films on silicon. The advancements in physical vapor deposition techniques have lead to notable progress in achieving heterogeneous integration of Ce:YIG thin films on silicon substrates in recent years[11, 15-18]. The Ce:YIG thin films were either epitaxially grown on garnet single crystal substrates and wafer-bonded to silicon, or directly deposited on silicon substrates. For epitaxial Ce:YIG thin films grown by sputtering, the Faraday rotation angle reached -4500 deg/cm. The propagation loss was 40 dB/cm when bonded to silicon waveguides. The material figure of metrit (FoM, Faraday rotation divided by the absorption per length) was 112.5 deg/dB[19, 20]. For Ce:YIG films deposited on silicon by pulsed laser deposition or sputtering, the Faraday rotation reached -5523 deg/cm at 1550 nm wavelength and the propagation loss was 72.7 dB/cm. The material FoM was 75.9 deg/dB.

Despite of the advances, wafer-scale heterogeneous integration of high quality, high uniformity Ce:YIG thin films on silicon has not been achieved. Previous studies have primarily focused on the deposition of magneto-optical thin films on garnet or silicon substrates with limited areas of ~$mm^2$ size, , limiting the development of large scale integrated magneto-photonic circuits [9, 21-27]. Moreover, the characterization of magneto-optical constants of Ce:YIG thin films is carried out primarily through free-space Faraday rotation and ellipticity hysteresis measurements. However, these approaches often require cleaving the substrate, thereby causing irreversible damage to it. Spectroscopic ellipsometry provides a non-destructive means for the characterization of MO garnet thin films [28-32]. Magneto-optical spectroscopic ellipsometry was also developed to measure the thickness, optical constants, and permittivity tensor of garnet thin films[30, 31]. However, wafer-scale magneto-ellipsometry characterization methodology for garnet thin films have not been developed.

Here, we report wafer-scale growth and wafer-scale nondestructive characterization of the full permittivity tensor of Ce:YIG thin film on silicon for the first time. Using glancing Using a confocal-sputtering set-up, we achieved uniform YIG seed layer deposition on 4 inch silicon using a 3 inch substrate, followed by uniform Ce:YIG layer deposition. A series of wafer-scale characterization techniques, including X-ray diffraction (XRD), electron backscatter diffraction (EBSD), X-ray photoelectron spectroscopy (XPS), and vibrating sample magnetometry (VSM), demonstrated high crystallinity, uniform valence state distribution and magnetic property the Ce:YIG film across the entire silicon wafer. A custom designed and assembled magneto-ellipsometer applied a transverse magnetic field over the focused light spot during spectroscopic ellipsometry measurements. Wafer-scale magneto-ellipsometer is custom designed and assembled by placing a pair of permanent magnet over the light focal spot. A uniform transverse magnetic field up to 578 Oe is applied under the Voigt geomery. Based on Mueller scattering matrix measurement results, model fitting and the transfer matrix method, the thickness, optical constants, and magneto-optical constants distribution were characterized non-destructively across the entire 4-inch silicon wafer. Ce:YIG thin films exhibiting thickness uniformity better than 3.5% and optical-constant uniformity better than 0.3% over the 4-inch wafer. A series of ring-resonator type optical isolators were fabricated along the wafer radial direction with comparable device performance, demonstrating the potential for large-scale magneto-photonic circuit integration.

## 2. Experimental Section/Methods

### Material Fabrication and Characterization

YIG and Ce:YIG thin films were deposited via confocal-magnetron sputtering using stoichiometric dielectric targets. The targets were prepared using $Y_2O_3$, $CeO_2$ and $Fe_2O_3$ powders (Sigma-Aldrich, 99.99%) and solid-state reaction method. Using a two-step deposition method, an approximately 60 nm thick YIG layer was first deposited on a 4-inch silicon wafer by radio-frequency magnetron sputtering (Leybold UNIVEX 400) and then crystallized by rapid thermal annealing to serve as the seed layer for the Ce:YIG film. Subsequently, a Ce:YIG layer with a thickness of approximately 120 nm was deposited on the YIG seed layer using the same method, followed by crystallization via rapid thermal annealing. To ensure the uniform thickness of the deposited film on the wafer, the wafer undergoes continuous rotation at a constant speed of 30 rpm during the sputtering process. A two-step magnetron sputtering deposition method is applied. First, an amorphous YIG layer with a thickness of 60 nm was deposited on the silicon wafer at room temperature. with Ar Rapid thermal annealing was performed at 850 ℃ for in oxygen to crystallized YIG. This step was followed by the deposition of a 120 nm thick layer of Ce:YIG on top of the YIG seed layer at room temperature. The resulting film, denoted as Ce:YIG/YIG/Si, then underwent rapid thermal annealing at 950 ℃ under an oxygen partial pressure of 1.5 Torr for 100 seconds to crystallize Ce:YIG.

The crystal structure was characterized using an X-ray diffractometer (Rigaku Ultima IV) equipped with a Cu-Kα radiation source. XRD patterns were collected over a 2θ range of 25°–40° with a step size of 0.02°, and the lattice constant of the Ce:YIG thin film was determined by fitting the XRD data. X-ray photoelectron spectroscopy (XPS) was measured on a Physical Electronics PHI Quantera Scanning X-ray Microprobe using mono-chromatic Al Ka radiation. The crystallized fraction was measured using an Oxford Instruments EBSD detector. Room temperature (300 K) magnetic hysteresis loops were measured using a vibrating sample magnetometry (VSM) model (Cryogenic, UK) equipped with a superconducting magnet.

SiN ring resonators were manufactured in a silicon photonics foundry. The SiN waveguides were formed on a $SiO_2$ lower cladding layer using LPCVD. After deposition of the upper $SiO_2$ cladding, CMP was applied to achieve planarization, and reactive ion etching (RIE) was then performed in the MO waveguide region to expose the SiN core. 6 SiN ring resonator type MO isolators were fabricated by placing the devices from the wafer center to the edge followed by MO thin films deposition. The transmission spectra were characterized using a polarization-maintaining, fiber butt-coupled measurement setup.

**Wafer-scale Magneto-Optical Ellipsometery**

We characterized the magneto-optical properties of the thin films across the wafer using a spectroscopic ellipsometer. Two permanent magnets were suspended above the measurement spot to apply a magnetic field to the films. During the measurements, the sample stage was translated to characterize the films at different positions across the wafer. Multiple points on the wafer is characterized for spectroscopic ellipsometer measurements. A total of 54 test points was chosen, and a scanning path was defined to sequentially measure the Muller matrix elements of the MO thin film at each test point, following a spiral pattern from the center towards the outer edge. The Muller matrix elements were measured under positive and negative magnetic field orientations. The ellipsometric parameters, ψ and Δ, were measured in reflection mode over a spectral range of 210 nm to 1690 nm (0.7eV to 5.9eV) at an incident angle of 70°. Applying of a transverse magnetic field, we measured the Mueller matrix of the MO thin film under different magnetic field directions. After each spectrum measurement the orientation of the magnetization was reversed. To reduce the effect of random noise or increase measurement sensitivity, we repeated the measurement five times. The permittivity tensor of Ce:YIG was then obtained by model fitting and the transfer-matrix method.

**Magneto-optical characterization of garnet films**

For an air/magneto-optical film/substrate structure, a linearly polarized light beam was obliquely incident on the film surface at a certain angle. An external magnetic field was applied parallel to the film surface and perpendicular to the plane of incidence. Under these conditions, the dielectric tensor of the magneto-optical material was expressed as follows:

$$\varepsilon=\begin{pmatrix} \varepsilon_1 & 0 & -i\cdot\varepsilon_2 \\ 0 & \varepsilon_1 & 0 \\ i\cdot\varepsilon_2 & 0 & \varepsilon_3 \end{pmatrix} \quad (1)$$

For linear magneto-optic effects, $\varepsilon_3 = \varepsilon_1$. $\varepsilon_1$ and $\varepsilon_2$ are :

$$\varepsilon_1 = \varepsilon_{1r} - i\cdot\varepsilon_{1i} \quad (2)$$

$$\varepsilon_2 = \varepsilon_{2r} - i\cdot\varepsilon_{2i} \quad (3)$$

$\varepsilon_1$ is first measured under 0 applied magnetic field. The ellipsometry data is fitted by the Tauc-Lorentz model, which defines the imaginary part of diagonal permittivity tensor elements as:

$$\varepsilon_{1i,TL}(E) = \frac{1}{E}\frac{AmpE_0Br\left(E-E_g\right)^2}{\left(E^2-E_0^2\right)^2+Br^2E},\left(E>E_g\right) \quad (4)$$

$$\varepsilon_{1i,TL}(E) = 0,\left(E\le E_g\right) \quad (5)$$

The real part of diagonal permittivity tensor elements can be obtained using analytical integration of K-K relation.

For ellipsometry measurements, we can obtain the ellipsometry parameters $\psi$, $\Delta$ and the $4\times4$ Mueller matrix, which can be used to describe the polarization changes of light reflected from the surface of a general sample. In the case of an isotropic material, the normalized reflection Mueller matrix is given in the block-diagonal form:

$$M=\begin{bmatrix} 1 & -N & 0 & 0 \\ -N & 1 & 0 & 0 \\ 0 & 0 & C & S \\ 0 & 0 & -S & C \end{bmatrix} \quad (6)$$

The Mueller matrix in Eq. (6) was scaled by $M_{11}$, which represented the overall reflected intensity under completely unpolarized illumination. For a nondepolarizing optical system, the parameters N, C and S were expressed in terms of the conventional ellipsometric angles $\Psi$ and $\Delta$.

$$N=\cos2\psi, C=\sin2\psi\sin\Delta, S=\sin2\psi\cos\Delta \quad (7)$$

The transfer matrix of the magneto-optical thin-film system was denoted by $Q=D_1^{-1}D_2P_2D_2^{-1}D_3$, from which the reflectances of s- and p-polarized components were calculated.

$$
\begin{aligned}
r_{ss} &= \frac{Q_{21}Q_{33} - Q_{23}Q_{31}}{Q_{11}Q_{33} - Q_{13}Q_{31}} \\
r_{sp} &= \frac{Q_{41}Q_{33} - Q_{43}Q_{31}}{Q_{11}Q_{33} - Q_{13}Q_{31}} \\
r_{ps} &= \frac{Q_{11}Q_{23} - Q_{21}Q_{13}}{Q_{11}Q_{33} - Q_{13}Q_{31}} \\
r_{pp} &= \frac{Q_{11}Q_{43} - Q_{41}Q_{13}}{Q_{11}Q_{33} - Q_{13}Q_{31}}
\end{aligned}
\tag{8}
$$

Meanwhile, the 4×4 Mueller matrix elements measured by the ellipsometer were also related to the reflectances, and their transformation relationship was expressed as follows:

$$
M = \begin{bmatrix}
\frac{1}{2}(|r_{pp}|^2 + |r_{sp}|^2 + |r_{ps}|^2 + |r_{ss}|^2) & \frac{1}{2}(|r_{pp}|^2 + |r_{sp}|^2 - |r_{ps}|^2 - |r_{ss}|^2) & \mathrm{Re}(r_{pp}r_{ps}^* + r_{sp}r_{ss}^*) & -\mathrm{Im}(r_{pp}r_{ps}^* + r_{sp}r_{ss}^*) \\
\frac{1}{2}(|r_{pp}|^2 - |r_{sp}|^2 + |r_{ps}|^2 - |r_{ss}|^2) & \frac{1}{2}(|r_{pp}|^2 - |r_{sp}|^2 - |r_{ps}|^2 + |r_{ss}|^2) & \mathrm{Re}(r_{pp}r_{ps}^* - r_{sp}r_{ss}^*) & -\mathrm{Im}(r_{pp}r_{ps}^* - r_{sp}r_{ss}^*) \\
\mathrm{Re}(r_{pp}r_{sp}^* + r_{ps}r_{ss}^*) & \mathrm{Re}(r_{pp}r_{sp}^* - r_{ps}r_{ss}^*) & \mathrm{Re}(r_{pp}r_{ss}^* + r_{ps}r_{sp}^*) & -\mathrm{Im}(r_{pp}r_{ss}^* - r_{ps}r_{sp}^*) \\
\mathrm{Im}(r_{pp}r_{sp}^* + r_{ps}r_{ss}^*) & \mathrm{Im}(r_{pp}r_{sp}^* - r_{ps}r_{ss}^*) & \mathrm{Im}(r_{pp}r_{ss}^* - r_{ps}r_{sp}^*) & \mathrm{Re}(r_{pp}r_{ss}^* + r_{ps}r_{sp}^*)
\end{bmatrix}
\tag{9}
$$

Accordingly, the off-diagonal Mueller matrix elements were fitted using the transfer-matrix method.

## 3. Results and Discussion

Figure 1A depicted the schematic representation of the confocal magnetron radio frequency sputtering system used for YIG and Ce:YIG deposition, showcasing the sputtering source, a 3-inch target, and the 4-inch silicon wafer within the growth chamber. The off-axis configuration allowed homogeneous deposition of Ce:YIG thin films on a 4-inch silicon wafer. The confocal setup also allowed YIG/Ce:YIG bilayer deposition.

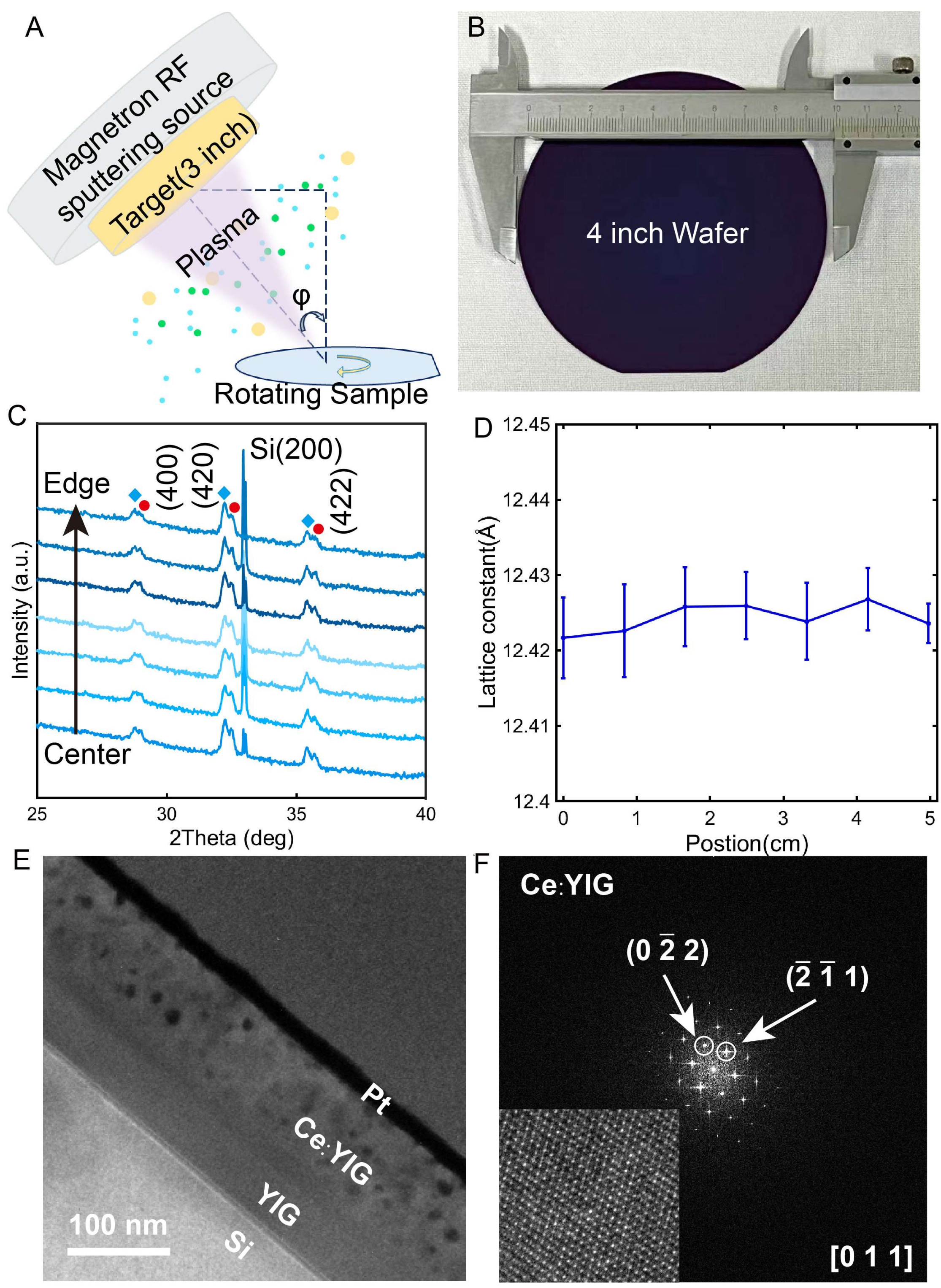

**Figure 1. (a)** Schematic diagram of sputtering deposition of 4-inch silicon-based MO thin films. **(b)** Comparison of substrate sizes used for the Ce:YIG thin-films growth. **(c)** XRD patterns of the Ce:YIG thin film on the wafer from the center to the edge. **(d)** Lattice Constants of CeYIG Films at Different Radial Positions on the Wafer **(e)** and **(f)** STEM images of the Ce:YIG thin film and the Fourier transform pattern of the lattice

Figure 1B showed a 4-inch magneto-optical thin film wafer. Detailed structural characterization was conducted to investigate the solubility of Ce in the YIG lattice, aiming to provide insights into the physical properties of this film. Figure 1C was the X-ray diffraction patterns of Ce:YIG thin film, showing well-crystallized diffraction peaks of the polycrystalline garnet phase. Due to the larger ionic radius of $Ce^{3+}$ (1.15 Å) compared to $Y^{3+}$ (1.02 Å), the lattice constant of Ce:YIG was larger than that of YIG[8, 17]. The XRD diffraction peaks of Ce:YIG showed no obvious shift from the center to the edge of the wafer, indicating that the variation in lattice constant is negligible. The calculated lattice constant was shown in Figure 1D. A bright-field TEM cross-sectional image of Ce:YIG/YIG film on Si(100) was shown in Figure 1E. While the Ce:YIG film maintains a homogeneous garnet phase in most area of the film, localized precipitation of $CeO_2$ nanocrystals with average diameters of 10.8 nm± 2.1 nm were observed. Figure 1F presented the high resolution lattice image and Fourier transform pattern of the Ce:YIG thin film along the [011] zone axis. The Fourier transform pattern reveals diffraction spots corresponding to the garnet crystal structure, with measured interplanar spacings of $d_{022}$ = 4.39 Å and $d_{211}$ = 5.07 Å, corresponding to lattice constant of 12.42 Å, a good match with XRD results.

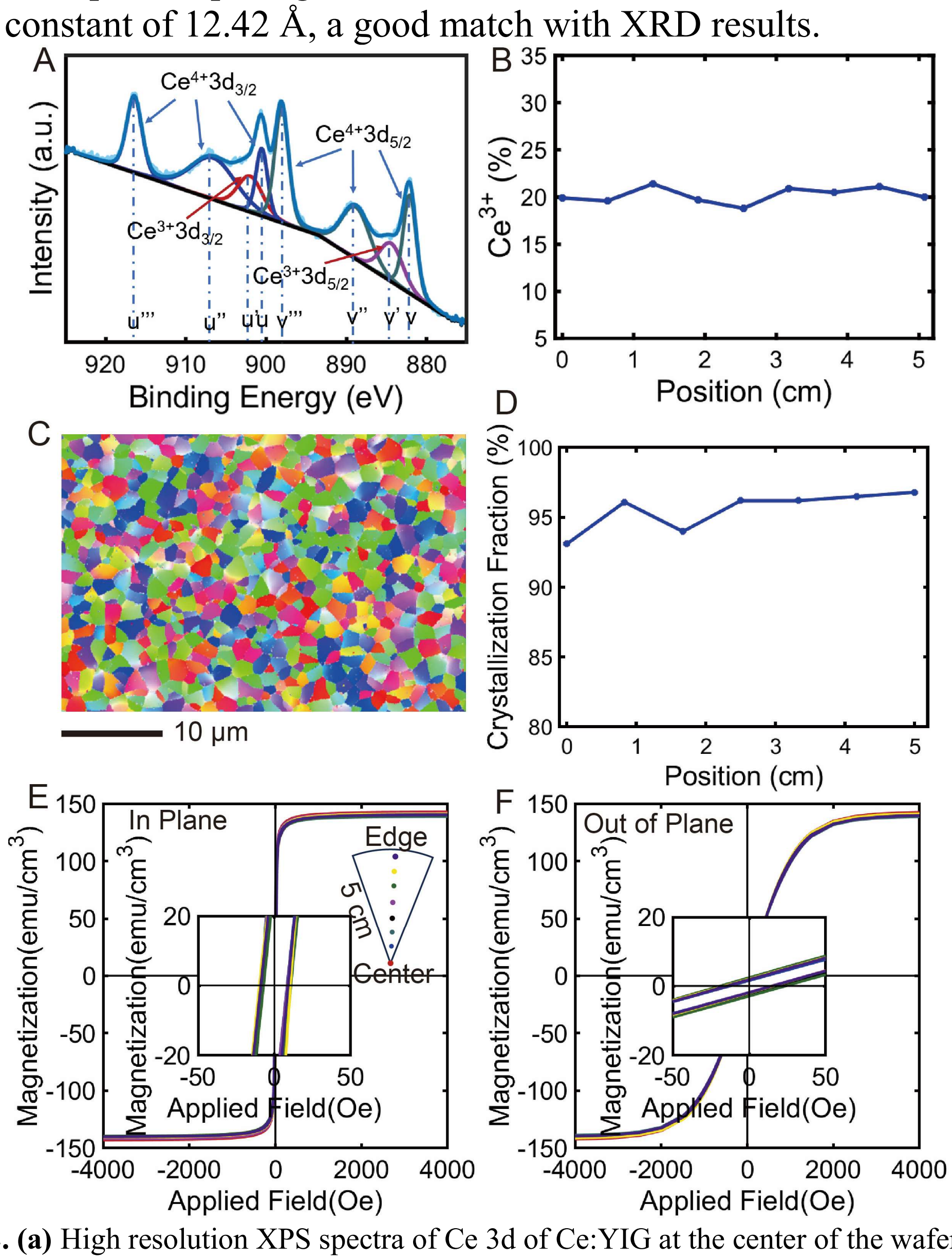

**Figure 2. (a)** High resolution XPS spectra of Ce 3d of Ce:YIG at the center of the wafer. **(b)** $Ce^{3+}$ content at different positions along the wafer radius. **(c)** EBSD map at the center of the wafer. **(d)** Crystallization fraction at different positions along the wafer radius. Room temperature **(e)** in-plane and **(f)** out of plane magnetic hysteresis of Ce:YIG films measured at different positions on the wafer.

To verify the uniformity of the 4-inch wafer-scale Ce:YIG thin film, a series of characterization were conducted at selected points along the radial direction of the wafer. Owing to the use of rotational deposition during the film growth process, sampling along the radial direction can, to a certain extent, serve as a representative approximation of the overall thin-film properties across the entire wafer. To quantitatively study the Ce valence states, XPS characterization was carried out. Figure 2A shows the XPS Ce 3d spectra of the Ce:YIG thin film at the center of the wafer. The C1s peak arised from hydrocarbon on the film surface at 284.8 eV was used as a reference for absolute binding energy calibration. The peaks assigned for $Ce^{4+}$ located at v (882.2 eV), v'' (888.7 eV), v''' (898.1 eV), u (900.5 eV), u'' (906.8 eV), u''' (916.5 eV) were attributed to $Ce^{4+}$. The v' (884.6 eV) and u' (902.3 eV) originate from the $3d_{5/2}$ and $3d_{3/2}$ peaks of the $Ce^{3+}$ ions. Figure 2B showed the $Ce^{3+}$ content at different positions along the wafer radius, indicating that the $Ce^{3+}$ content remains around 20% across the wafer. Figures 2C and 2D showed the EBSD map at the wafer center and the crystallized fraction of Ce:YIG along the wafer radius, respectively. Ce:YIG films maintained a high crystallization fraction of over 93% at different locations across the wafer, indicating the film was predominately garnet phase.

As shown in Figure 2E and 2F, we measured the room-temperature magnetic hysteresis loops of Ce:YIG thin films at different positions on the wafer. The saturation magnetization values of the Ce:YIG films at different positions were all within $143.5 \pm 1.5$ emu/cm³, which was consistent with values reported in other literature[17]. Considering the volume error in the measured films, the saturation magnetization of the Ce:YIG films at different positions on the wafer were almost identical. All these films exhibited easy plane magnetic anisotropy. The in-plane saturation magnetic field was approximately 525 Oe, with a coercivity of $9.5\pm0.7$ Oe, while the out-of-plane saturation magnetic field was around 2250 Oe, with a coercivity of $16.6\pm0.9$ Oe.

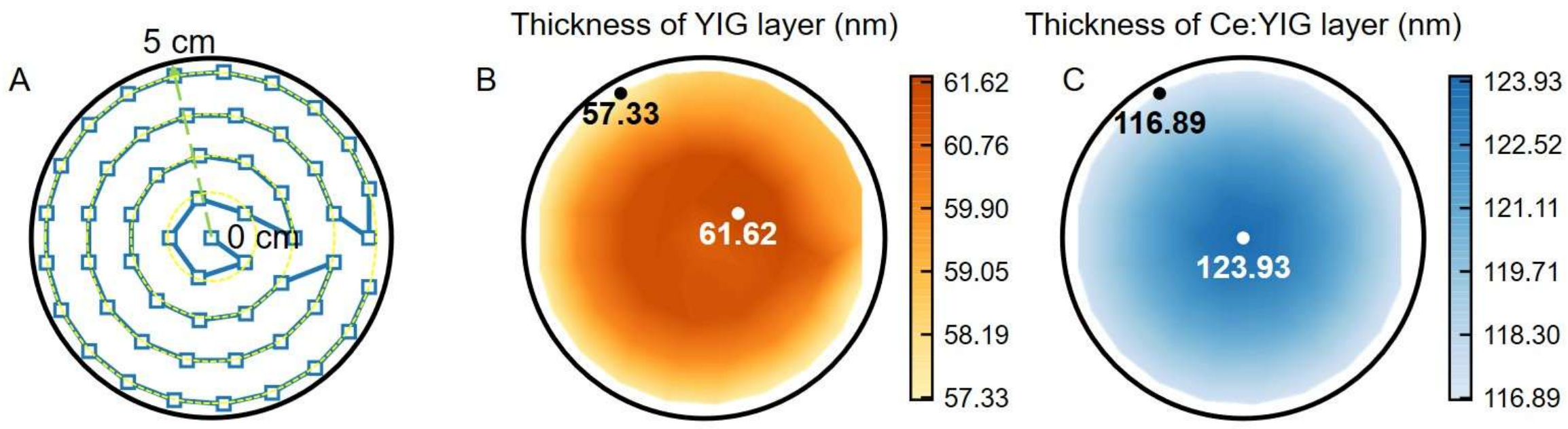

**Figure 3. (a)** The scanning trajectory of the magneto-ellipsometry measurement. **(b)** Measured thickness distribution of YIGand **(c)** Measured thickness distribution of Ce:YIG

To measure the thickness of wafer-scale YIG and Ce:YIG thin films, an automatic scanning program was set up across the whole wafer. Figure 3A showed the scanning trajectory of the ellipsometery measurement across the wafer. The squares in the schematic represented the probing locations on the wafer. By fitting the data, the thicknesses of the YIG and Ce:YIG films can be obtained. As depicted in Figure 3B and 3C, the thickness of the Ce:YIG layer was determined to be approximately 120 nm with a variation ratio of around 3.5% across the wafer. Meanwhile, the thickness of the YIG layer is approximately 60 nm, with a variation ratio of about 3.3% across the wafer. Additionally, both the Ce:YIG layer and YIG layer films exhibited a radial downward trend in thickness from the center to the edge, which can be attributed to the inevitable error produced by magnetron sputtering when processing large-area wafers. For confocal sputtering, it was relatively difficult to deposit thin films with in-plane thickness uniformity over large-area substrates[33, 34].

In order to characterize the optical and MO properties of this film, we measured the Mueller matrix of the magneto-optical thin films using spectroscopic ellipsometry. Meanwhile, we used the transfer matrix method to calculate the real and imaginary parts of the off-diagonal elements of the Ce:YIG thin film.

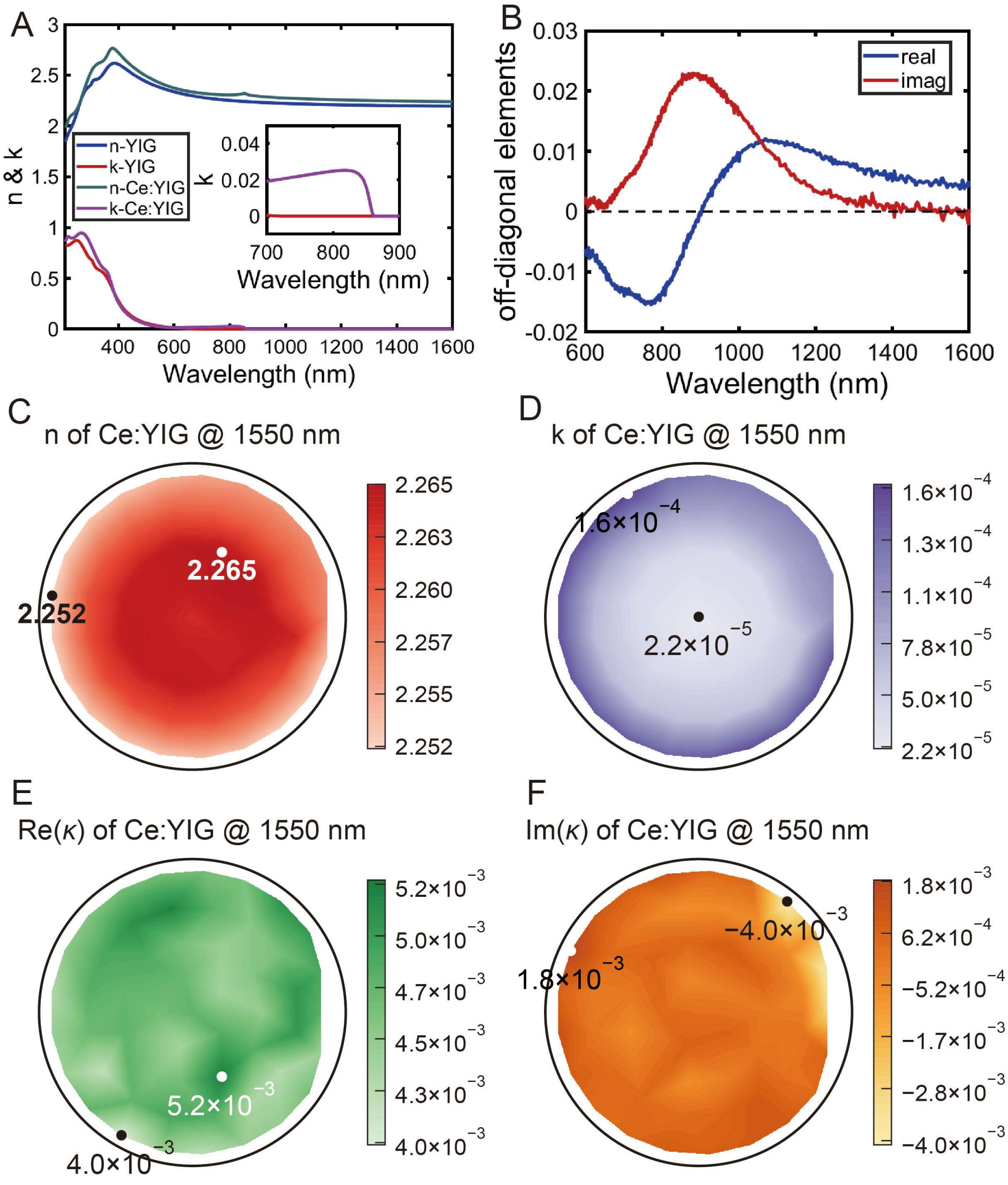

**Figure 4. (a)** The spectra of the refractive index (n) and extinction coefficient (k) of YIG and CeYIG at the center of the wafer. **(b)** The spectra of the real and imaginary parts of the off-diagonal elements

of the dielectric tensor of the CeYIG thin film at the center of the wafer. **(c)** and **(d)** Map of the optical constant n and k of YIG thin film at 1550 nm on 4-inch silicon. **(e)** and **(f)** Map of the real part and imaginary part of off-diagonal permittivity tensor elements of Ce: YIG thin films at 1550 nm on 4-inch silicon.

Figure 4A showed the derived spectra of optical constants n and k of the YIG layer and Ce:YIG layer. Owing to the incorporation of Ce dopants, the refractive index of Ce:YIG is generally higher than that of YIG thin films over the entire wavelength range. In contrast, the extinction coefficient of the Ce:YIG thin film exhibited an absorption peak around 832 nm. This absorption arised from the intra-ionic electrical dipole transitions of the $Ce^{3+}$ ions[31]. Figure 4B exhibited the real and imaginary parts of off-diagonal permittivity tensor elements in the MO thin film Ce:YIG, which were obtained by transfer matrix method. For the real part of off-diagonal permittivity tensor elements, we observed global maxima at 800 nm. This is the result of the *4f-5d* transition in $Ce^{3+}$ ions at 800 nm[35]. At the wavelength of 1550 nm, the real part of off-diagonal permittivity tensor elements of the Ce:YIG MO film was determined to be approximately 0.0046, with a variation ratio of around 13% across the wafer. The error may be caused by jitter-induced noise from the light source. In the future, we can further reduce the noise by increasing the light-source intensity. For the imaginary part of off-diagonal permittivity tensor elements, we observed global maxima at 880 nm, which is also related to an intra-ionic electrical dipole transition of $Ce^{3+}$[31]. At the wavelength of 1550 nm, the imaginary part of off-diagonal permittivity tensor elements was also found to be negligibly small and can be regarded as effectively zero.

For the 4-inch wafer, as shown in Figure 4C, the refractive index of Ce:YIG exhibits a distribution characterized by higher values at the center and lower values toward the edge. In contrast, the extinction coefficient showed the opposite trend, with lower values at the center and higher values near the edge. The refractive index n of the YIG MO film was determined to be approximately 2.218 $\pm 0.006$, with a variation ratio of around 0.25% across the wafer. Additionally, the refractive index n of the Ce:YIG MO film was determined to be $2.258 \pm 0.007$, with a variation ratio of around 0.30% across the wafer. This behavior was likely attributed to subtle variations in processing conditions across the wafer during the sputtering deposition and annealing processes. Figure 4D showed the wafer-scale distribution of the extinction coefficient k of the top Ce:YIG film at a wavelength of 1550 nm. The maximum k value of the Ce:YIG film was only $1.63 \times 10^{-4}$, while the minimum was as low as $2.20 \times 10^{-5}$. In the wavelength of 1550 nm, where optical absorption was extremely weak, the reliability of the fitting results was relatively low, as the measurement has approached the limit of the instrument's accuracy. Additional experimental verification was therefore needed to obtain more reasonable and accurate results. Figures 4E and 4F showed the wafer-scale distributions of the off-diagonal elements of the permittivity tensor of the Ce:YIG film. At the wavelength of 1550 nm, both the real and imaginary parts of the off-diagonal dielectric tensor elements of the Ce:YIG film were relatively small, making the measurements more susceptible to intensity noise of the ellipsometer light source. As a result, the measured overall wafer-scale variation is slightly higher, at approximately 9%. The relatively large wafer-scale variation in the off-diagonal elements of the Ce:YIG dielectric tensor may arise not only from the strong influence of measurement noise, but also from the stepwise amplification of differences in lattice constant, valence state, and optical constants during the transfer-matrix calculation process, which ultimately leads to a larger apparent wafer-scale variation. These results demonstrated the feasibility of non-destructive characterization of wafer-scale magneto-optical properties using magneto-ellipsometry wafer mapping combined with the transfer-matrix method.

To demonstrate the capability for wafer-scale fabrication of integrated optical isolators, six microring devices were positioned at different radial locations on the silicon wafer, spanning from the center to a radial distance of 5 cm. Figure 5A showed the spatial distribution of the six devices across the wafer. An optical microscope image of a TM-mode optical isolator was presented in Figure 5B, where the isolator is based on a SiN racetrack resonator structure. In addition, Figure 5C provided the simulated $E_y$ field distribution of the fundamental TM mode of the MO/SiN waveguide. These devices employ a strip-loaded waveguide configuration, consisting of a planar stack of Ce:YIG (97 nm)/YIG (50 nm) thin films deposited on the SiN layer. The variations in YIG thickness, Ce:YIG thickness, and refractive index among the optical isolators at different locations are minimal. Detailed simulations are provided in the Supplementary Information.

The SiN racetrack resonator was characterized using a polarization-maintaining, fiber-butt-coupled system. Across the wafer, the optical isolators located at different positions exhibit isolation ratios ranging from 18.6 to 31 dB and insertion losses between 2.8 and 4.3 dB. This demonstrated our capability for the large-scale fabrication of optical isolators. After the calculations, the Faraday rotation of the wafer-scale CeYIG thin-film material was between 2216 deg/cm and 2420 deg/cm, and the optical loss of the thin-film material was between 69 dB/cm and 91 dB/cm. However, there are still some defects in the CeYIG thin films, such as grain-boundary precipitates and $CeO_2$ precipitates inside the grains, which lead to lower Faraday rotation and higher optical loss. In the future, we will try to address these issues by tuning the thin-film composition and controlling the oxygen partial pressure during annealing.

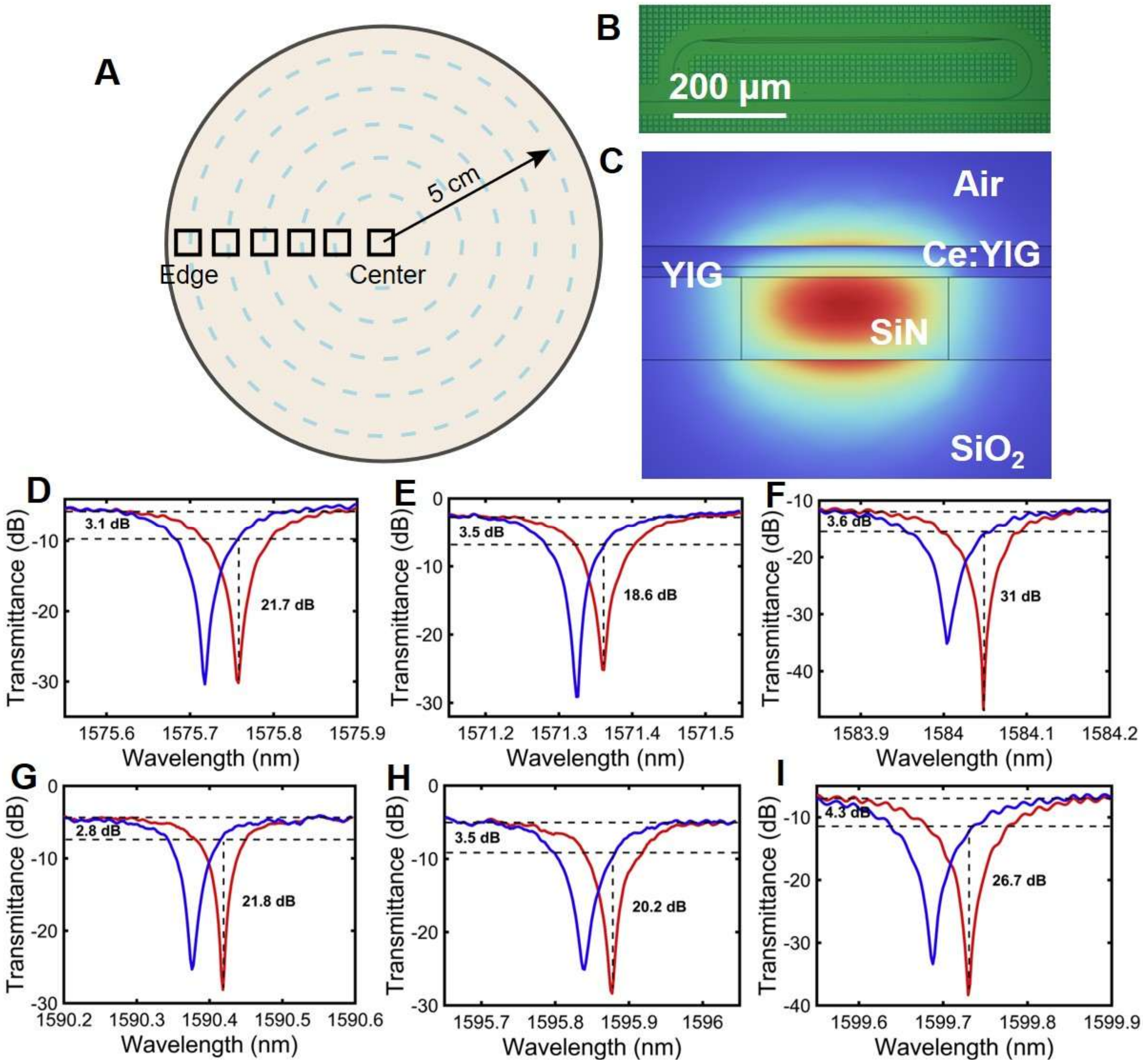

**Figure 5.** (**a**) Schematic illustration of the spatial distribution of optical isolators across the wafer. (**b**) Optical microscope image of the TM-mode optical isolator. **(c)** Simulated $E_y$ field distribution of the fundamental TM mode of the MO/SiN waveguide. **(d) - (i)** Forward and backward transmission spectra of optical isolators at different locations across the wafer.

With wafer-scale fabrication of magneto-optical thin films, we can envision system-level nonreciprocal photonic integrated circuits from scalable circulator or isolator arrays for dense optical communication, to nonreciprocal optical routing elements that act as fundamental building blocks for optical computing or interconnects. This technology is expected to enable wafer-scale fabrication of nonreciprocal devices, which can significantly reduce manufacturing costs.

## 4. Conclusion

In this study, magneto-optical Ce:YIG thin films were successfully grown on 4-inch silicon substrates using confocal magnetron sputtering. A wafer-scale non-destructive magneto-ellipsometry characterization method was setup to measure the thickness and full permittivity distribution across the whole wafer.

The fabricated film showed large Faraday rotation of 2318 ± 102 deg/cm, propagation loss of 80±11 dB/cm, thickness homogenity of 3.5%, optical constant variation of 2.8% and off-diagonal permittivity element variation of 13%. Microring-based optical isolator devices fabricated based on the sputter deposited films showed isolation ratio of 24.8±6.2 dB, insertion loss of 3.55±0.75 dB. Our study demosntrate a step forward toward wafer-scale integration of MO thin films on silicon, enabling high density integration of magneto-optical isolators, circulators, switches and modulators for optical communication, sensing and computation applications.

**Supporting Information**

Supporting Information is available from the Wiley Online Library or from the author.

## Acknowledgements

The authors are grateful for the support by the National Natural Science Foundation of China (NSFC) (Grant Nos. 52450018, U22A20148, 52021001 and 52473292), Sichuan Provincial Science and Technology Department (Grant Nos. 2025ZYD0001， 2025ZNSFSC0040, 2024NSFSC0484), the Open-Foundation of Key Laboratory of Laser Device Technology, China North Industries Group Corporation Limited (Grant No. K20032-5)

## References

[1] W. Yan *et al.*, "Waveguide-integrated high-performance magneto-optical isolators and circulators on silicon nitride platforms," *Optica,* vol. 7, no. 11, pp. 1555-1562, 2020/11/20 2020, doi: 10.1364/OPTICA.408458.

[2] W. Yan *et al.*, "Nonreciprocal optical routing via a magneto-optical phased array on silicon," *Photon. Res.,* vol. 13, no. 9, pp. 2432-2441, 2025/09/01 2025, doi: 10.1364/PRJ.547240.

[3] P. Pintus *et al.*, "Integrated non-reciprocal magneto-optics with ultra-high endurance for photonic in-memory computing," *Nature Photonics,* vol. 19, no. 1, pp. 54-62, 2025/01/01 2025, doi: 10.1038/s41566-024-01549-1.

[4] P. Pintus *et al.*, "An integrated magneto-optic modulator for cryogenic applications," *Nature Electronics,* vol. 5, no. 9, pp. 604-610, 2022/09/01 2022, doi: 10.1038/s41928-022-00823-w.

[5] R. Hisatomi *et al.*, "Bidirectional conversion between microwave and light via ferromagnetic magnons," *Physical Review B,* vol. 93, no. 17, p. 174427, 05/27/ 2016, doi: 10.1103/PhysRevB.93.174427.

[6] W. Yan *et al.*, "Ultra-broadband magneto-optical isolators and circulators on a silicon nitride photonics platform," *Optica,* vol. 11, no. 3, pp. 376-384, 2024/03/20 2024, doi: 10.1364/OPTICA.506366.

[7] W. Yan *et al.*, "Waveguide-integrated high-performance magneto-optical isolators and circulators on silicon nitride platforms," *Optica,* vol. 7, no. 11, 2020, doi: 10.1364/optica.408458.

[8] Y. Zhang *et al.*, "Enhanced magneto-optical effect in Y1.5Ce1.5Fe5O12 thin films deposited on silicon by pulsed laser deposition," *Journal of Alloys and Compounds,* vol. 703, pp. 591-599, 2017/05/05/ 2017, doi: https://doi.org/10.1016/j.jallcom.2017.01.315.

[9] Q. Du, T. Fakhrul, Y. Zhang, J. Hu, and C. A. Ross, "Monolithic magneto-optical oxide thin films for on-chip optical isolation," *MRS Bulletin,* vol. 43, no. 6, pp. 413-418, 2018, doi: 10.1557/mrs.2018.127.

[10] Y. Yoshihara *et al.*, "Thickness-dependent magnetooptical properties of ion beam sputtered polycrystalline Ce1Y2Fe5O12 films," *Optical Materials,* vol. 133, p. 112967, 2022/11/01/ 2022, doi: https://doi.org/10.1016/j.optmat.2022.112967.

[11] S. Ghosh, S. Keyvavinia, W. Van Roy, T. Mizumoto, G. Roelkens, and R. Baets, "Ce:YIG/Silicon-on-Insulator waveguide optical isolator realized by adhesive bonding," *Opt. Express,* vol. 20, no. 2, pp. 1839-1848, 2012/01/16 2012, doi: 10.1364/OE.20.001839.

[12] Y. Zhang *et al.*, "Monolithic integration of broadband optical isolators for polarization-diverse silicon photonics," *Optica,* vol. 6, no. 4, pp. 473-478, 2019/04/20 2019, doi: 10.1364/OPTICA.6.000473.

[13] Y. Shoji and T. Mizumoto, "Silicon Waveguide Optical Isolator with Directly Bonded Magneto-Optical Garnet," *Applied Sciences*, vol. 9, no. 3*,* p. 609doi: 10.3390/app9030609.

[14] P. Mihailovic and S. Petricevic, "Fiber Optic Sensors Based on the Faraday Effect," *Sensors*, vol. 21, no. 19*,* p. 6564doi: 10.3390/s21196564.

[15] K. Srinivasan *et al.*, "High-Gyrotropy Seedlayer-Free Ce:TbIG for Monolithic Laser-Matched SOI Optical Isolators," *ACS Photonics,* vol. 6, no. 10, pp. 2455-2461, 2019/10/16 2019, doi: 10.1021/acsphotonics.9b00707.

[16] K. Srinivasan, N. C. A. Seaton, R. Peng, M. Li, and B. J. H. Stadler, "Crystallization of high gyrotropy garnets with decreasing thermal processing budgets as analyzed by electron backscatter diffraction," *Opt. Mater. Express,* vol. 13, no. 2, pp. 357-367, 2023/02/01 2023, doi: 10.1364/OME.476482.

[17] Y. Yang *et al.*, "Europium-substituted cerium iron garnet thin films for silicon-integrated nonreciprocal photonic device applications," *APL Materials,* vol. 13, no. 5, p. 051109, 2025, doi: 10.1063/5.0256931.

[18] Y. Shoji, M. Ito, Y. Shirato, and T. Mizumoto, "MZI optical isolator with Si-wire waveguides by surface-activated direct bonding," *Opt. Express,* vol. 20, no. 16, pp. 18440-18448, 2012/07/30 2012, doi: 10.1364/OE.20.018440.

[19] Y. Shoji, T. Mizumoto, H. Yokoi, I. W. Hsieh, and R. M. Osgood, Jr., "Magneto-optical isolator with silicon waveguides fabricated by direct bonding," *Applied Physics Letters,* vol. 92, no. 7, p. 071117, 2008, doi: 10.1063/1.2884855.

[20] S. Liu, D. Minemura, and Y. Shoji, "Silicon-based integrated polarization-independent magneto-optical isolator," *Optica,* vol. 10, no. 3, pp. 373-378, 2023/03/20 2023, doi: 10.1364/OPTICA.483017.

[21] T. Shintaku, A. Tate, and S. Mino, "Ce-substituted yttrium iron garnet films prepared on Gd3Sc2Ga3O12 garnet substrates by sputter epitaxy," *Applied Physics Letters,* vol. 71, no. 12, pp. 1640-1642, 1997, doi: 10.1063/1.120003.

[22] K. Hyonju, A. M. Grishin, K. V. Rao, S. C. Yu, R. Sbiaa, and H. L. Gall, "Ce-substituted YIG films grown by pulsed laser deposition for magneto-optic waveguide devices," *IEEE Transactions on Magnetics,* vol. 35, no. 5, pp. 3163-3165, 1999, doi: 10.1109/20.801115.

[23] B. Lei, H. Juejun, F. D. Gerald, K. Lionel, and C. A. Ross, "Monolithic integration of chalcogenide glass/iron garnet waveguides and resonators for on-chip nonreciprocal photonic devices," in *Proc.SPIE*, 2011, vol. 7941, p. 794105, doi: 10.1117/12.875184. [Online]. Available: https://doi.org/10.1117/12.875184

[24] T. Goto, M. C. Onbaşlı, and C. A. Ross, "Magneto-optical properties of cerium substituted yttrium iron garnet films with reduced thermal budget for monolithic photonic integrated circuits," *Opt. Express,* vol. 20, no. 27, pp. 28507-28517, 2012/12/17 2012, doi: 10.1364/OE.20.028507.

[25] S. Liu, Y. Shoji, and T. Mizumoto, "TE-mode magneto-optical isolator based on an asymmetric microring resonator under a unidirectional magnetic field," *Opt. Express,* vol. 30, no. 6, pp. 9934-9943, 2022/03/14 2022, doi: 10.1364/OE.454751.

[26] S. Liu, Y. Shoji, and T. Mizumoto, "Mode-evolution-based TE mode magneto-optical isolator using asymmetric adiabatic tapered waveguides," *Opt. Express,* vol. 29, no. 15, pp. 22838-22846, 2021/07/19 2021, doi: 10.1364/OE.427914.

[27] A. D. Block, P. Dulal, B. J. H. Stadler, and N. C. A. Seaton, "Growth Parameters of Fully Crystallized YIG, Bi:YIG, and Ce:YIG Films With High Faraday Rotations," *IEEE Photonics Journal,* vol. 6, no. 1, pp. 1-8, 2014, doi: 10.1109/JPHOT.2013.2293610.

[28] D. Rosu, P. Petrik, G. Rattmann, M. Schellenberger, U. Beck, and A. Hertwig, "Optical characterization of patterned thin films," *Thin Solid Films,* vol. 571, pp. 601-604, 2014/11/28/ 2014, doi: https://doi.org/10.1016/j.tsf.2013.11.052.

[29] M. A. Razooqi Alaani *et al.*, "Tailoring the CdS/CdSe/CdTe multilayer structure for optimization of photovoltaic device performance guided by mapping spectroscopic ellipsometry," *Solar Energy Materials and Solar Cells,* vol. 221, p. 110907, 2021/03/01/ 2021, doi: https://doi.org/10.1016/j.solmat.2020.110907.

[30] E. Jesenska *et al.*, "Optical and magneto-optical properties of Bi substituted yttrium iron garnets prepared by metal organic decomposition," *Opt. Mater. Express,* vol. 6, no. 6, pp. 1986-1997, 2016/06/01 2016, doi: 10.1364/OME.6.001986.

[31] M. C. Onbasli *et al.*, "Optical and magneto-optical behavior of Cerium Yttrium Iron Garnet thin films at wavelengths of 200–1770 nm," *Scientific Reports,* vol. 6, no. 1, p. 23640, 2016/03/30 2016, doi: 10.1038/srep23640.

[32] L. Halagačka *et al.*, "Mueller matrix optical and magneto-optical characterization of Bi-substituted gadolinium iron garnet for application in magnetoplasmonic structures," *Opt. Mater. Express,* vol. 4, no. 9, pp. 1903-1919, 2014/09/01 2014, doi: 10.1364/OME.4.001903.

[33] C. Z. Jiang, J. Q. Zhu, J. C. Han, P. Lei, and X. B. Yin, "Uniform film in large areas deposited by magnetron sputtering with a small target," *Surface and Coatings Technology,* vol. 229, pp. 222-225, 2013/08/25/ 2013, doi: https://doi.org/10.1016/j.surfcoat.2012.03.075.

[34] B. Nathan *et al.*, "Thickness distribution of sputtered films on curved substrates for adjustable x-ray optics," *Journal of Astronomical Telescopes, Instruments, and Systems,* vol. 5, no. 2, p. 021005, 3/1 2019, doi: 10.1117/1.JATIS.5.2.021005.

[35] M. Gomi, H. Furuyama, and M. Abe, "Strong magneto‐optical enhancement in highly Ce‐substituted iron garnet films prepared by sputtering," *Journal of Applied Physics,* vol. 70, no. 11, pp. 7065-7067, 1991, doi: 10.1063/1.349786.